\documentclass[a4paper,fleqn]{cas-sc}

\usepackage[authoryear]{natbib}
\usepackage{subcaption}
\usepackage[ascii]{inputenc}
\usepackage{graphicx}
\usepackage{subcaption}
 \usepackage{setspace}
\usepackage{textcomp}     
\usepackage{url}
\usepackage[]{amsmath}
\usepackage{amsmath}
\usepackage{nccmath}
\usepackage{lineno}

\newcommand{\vdy}{{1999 VD$_{57}$ }}

\newcommand{\vd}{{Dinkinesh }}
\newcommand{\vdns}{{Dinkinesh}}
\newcommand{\lc}{{\textit{Lucy} }}
\newcommand{\lcns}{{\textit{Lucy}}}
\newcommand*{\arcsec}{$^{\prime\prime}\mkern-1.2mu$}

\begin{document}
\let\WriteBookmarks\relax
\def\floatpagepagefraction{1}
\def\textpagefraction{.001}

\shorttitle{Spectral characterization of (152830) Dinkinesh}

\shortauthors{Bolin et al.}  

\title[mode = title]{Keck and Gemini spectral characterization of \textit{Lucy} mission fly-by target (152830) Dinkinesh}
\author[]{B. T. Bolin$^{a,*,**}$}[orcid=0000-0002-4950-6323]
\author[1]{K. S. Noll}[orcid=0000-0002-6013-9384]
\author[2]{I. Caiazzo}[orcid=0000-0002-4770-5388]
\author[3]{C. Fremling}[orcid=0000-0002-4223-103X]
\author[4]{R. P. Binzel}[orcid=0000-0002-9995-7341]
\address[1]{Goddard Space Flight Center, 8800 Greenbelt Road, Greenbelt, MD 20771, USA}
\address[2]{Division of Physics, Mathematics and Astronomy, California Institute of Technology, Pasadena, CA 91125, USA}
\address[3]{Caltech Optical Observatory, California Institute of Technology, Pasadena, CA 91125, USA}
\address[4]{Department of Earth, Atmospheric, and Planetary Sciences, Massachusetts Institute of Technology, Cambridge, MA 02139, USA}
\cortext[cor1]{Corresponding author: bryce.bolin@nasa.gov}
\cortext[cor2]{NASA Postdoctoral Program Fellow.} 

\begin{abstract}[S U M M A R Y]
Recently, the inner main belt asteroid (152830) Dinkinesh was identified as an additional fly-by target for the \lc mission. The heliocentric orbit and approximate absolute magnitude of \vd are known, but little additional information was available prior to its selection as a target. In particular, the lack of color spectrophotometry or spectra made it impossible to assign a spectral type to \vd from which its albedo could be estimated. We set out to remedy this knowledge gap by obtaining visible wavelength spectra with the Keck telescope on 2022 November 23 and with Gemini-South on 2022 December 27. The spectra measured with the Keck I/Low Resolution Imaging Spectrometer (LRIS) and the Gemini South/Gemini Multi-Object Spectrograph South (GMOS-S) are most similar to the average spectrum of S- and Sq-type asteroids. The most diagnostic feature is the  $\approx$15$\pm$1$\%$ silicate absorption feature at $\approx$0.9-1.0~micron. Small S- and Sq-type asteroids have moderately high albedos ranging from 0.17-0.35. Using this albedo range for \vd in combination with measured absolute magnitude, it is possible to derive an effective diameter and surface brightness for this body. The albedo, size and surface brightness are important inputs required for planning a successful encounter by the \lc spacecraft.\end{abstract}

\maketitle
\section{Introduction}
The \lc mission will be the first spacecraft mission to encounter members of the Jupiter Trojan asteroid population, flying by a total of eight Trojan objects from 2027 through 2033 \citep{Levison2021Lucy,Olkin2021Lucy}. In addition to Trojans, \lc is also planned to encounter the main belt asteroid (52246) Donaldjohanson in April 2025. A search of smaller asteroids that would come close to \lcns's trajectory recently identified one with an exceptionally close encounter in November 2023 (R.~Marschall, private communication). (152830) \vdy, recently named Dinkinesh \citep{Ticha2023}, orbits in the inner main belt with a semi-major axis of $a$  = 2.19 AU, eccentricity, $e$ = 0.11 and inclination $i$ = 2.09 degrees\footnote{\url{https://ssd.jpl.nasa.gov/tools/sbdb_lookup.html\#/?sstr=1999\%20VD57}}. With the addition of a small maneuver, \lc will be redirected to a $\approx$ 450 km distant flyby encounter with \vd on 01 November 2023\footnote{\url{https://www.nasa.gov/feature/goddard/2023/nasa-s-lucy-team-announces-new-asteroid-target}}, making it the first of ten asteroid targets that the spacecraft will visit.

A target's diameter is needed to plan an encounter with the \lc spacecraft. In order to determine an effective diameter $\mathrm{D_{eff}}$ from the measured absolute magnitude of the asteroid, $H$, it is necessary to estimate the visual geometric albedo. $\mathrm{D_{eff}}$ is proportional to $\mathrm{{p_v}^{-1/2}}$ and the size in km is given by ${\mathrm{D_{eff} = {1329}\ {{p_v}^{-1/2}} \ 10 ^{-{H/5}}}}$ \citep{Harris2002}. With no additional constraints, the possible albedo, $\mathrm{p_v}$, for an asteroid in the main belt spans a wide range, $0.04 \lesssim \mathrm{p_v} \lesssim 0.45$ \citep{DeMeo2013}, and the corresponding range of $\mathrm{D_{eff}}$ varies by a factor of more than three. 
 
 Both the diameter and albedo of \vd have implications for target acquisition and observation by the \lc spacecraft. \lc relies on an autonomous terminal tracking system to point the instrument platform at the target during the close approach sequence \citep{Olkin2021Lucy}. In the case of low albedo, the larger angular size will make earlier resolved imaging and acquisition with the terminal tracking system possible, increasing the time available to converge on an accurate range to the target. But low albedo could also make optical navigation at high phase angles in the weeks leading up to the close encounter more difficult due to the steeper phase function of lower albedo asteroids \citep{Pravec2012}. Alternatively, a higher albedo for \vd will have a smaller angular size at a given distance, reaching the terminal tracking system's threshold later in the sequence with less time to lock onto the target. The albedo may also impact observations of the resolved surface by rendering the target either too bright or too faint for the dynamic range of the instruments for a given exposure time.

Asteroid diameters are often derived through a combination of visible reflectance and thermal infrared emission \citep{Delbo2015}. More direct methods include radar observations \citep{Benner2015} or {\it in-situ} spacecraft observations \citep[e.g.,][]{Lauretta2017orex}. Polarimetry of asteroids taken at varying phase angles can also be used to infer albedos and diameters \citep{Belskaya2015}. For the case of \vdns, there are, unfortunately, no existing data that can be used to constrain its diameter. It remains possible, however, to use an indirect method for constraining albedo by exploiting known correlations of albedo and spectral type \citep{Thomas2011,DeMeo2013}. Therefore, we set out to measure the visible spectrum of \vd with the goal of determining the spectral-type and ultimately using that information to constrain its albedo and diameter.

\section{Observations}
We obtained two separate spectra, one from the Keck telescope and one from Gemini-South. We used the Low Resolution Imaging Spectrometer (LRIS) \citep[][]{Oke1995} on the Keck I 10 m telescope to observe \vd on 2022 November 23 in spectroscopy mode (Program ID C262, PI I. Caiazzo). We applied a similar observational strategy as described by \citep[][]{Bolin2020CD3, Bolin2021LD2,Bolin2022}. LRIS has two cameras - the blue camera consists of two 2k $\times$ 4k Marconi CCD arrays and the red camera consists of two science grade Lawrence Berkeley National Laboratory 2k $\times$ 4k CCD arrays. Both cameras have a spatial resolution of 0.135 arcsec pixel$^{-1}$. The 1.0-arcsecond wide slit was used with the 0.56~micron dichroic with $\approx$50$\%$ transmission efficiency in combination with the 600/4000 grating for the blue camera and the 400/8500 grating for the red camera providing a spectral resolution of 0.4~nm and 0.7~nm, respectively \citep[][]{Oke1995,McCarthy1998}. One 1200 s exposure and two 900 s exposures were taken of \vd for a total integration time of 3000 s. The exposures were taken in seeing conditions of $\approx$0.7\arcsec~measured at zenith and the observations were taken at an airmass of $\approx$1.05. Wavelength calibration used HgCdZn lamps for the blue camera and the ArNeXe lamps for the red camera. Flux calibration used the standard star BD+28 4211 for the blue and red camera, and a solar analog star, 2MASS HD53991, was used for slope correction. The spectra were reduced using the LPipe software for reducing LRIS data \citep[][]{Perley2019}.

On 2022 December 27, an additional spectrum of \vd was obtained using the Gemini-South Multi-Object Spectrograph (GMOS-S) on the Gemini South 8.1 m telescope was used to observe \vd  under program GS-2022B-FT-110 (PI: K. Noll). The GMOS-S detector array consists of three 2048 $\times$ 4176 Hamamatsu chips separated by 61-pixel gaps with a pixel scale of 0.08\arcsec pixel$^{-1}$ \citep[][]{Gimeno2016}. The 1.0-arcsecond wide slit was used with the R150 grating and the GG455$\_$G0329 order-blocking filter providing a spectral resolution of 1.2~nm \citep[][]{Hook2004}. Three 900 s exposures of \vd were taken for a total integration time of 2700 s and were taken in $\approx$0.7\arcsec~ seeing measured at zenith. Due to the +27$^\circ$ declination of \vd on 2022 December 27 and the $\approx$-30$^\circ$ latitude of the Gemini S site, the asteroid had to be observed when at the low elevation of $\approx$33$^\circ$ when it could be observed at a minimum airmass of $\approx$1.8 at $\approx$5:04 UTC on 2022 December 27. Observations of \vd closer to zenith could have been performed with a similar instrument at the Gemini North site on Maunakea, however, operations at Gemini North were not possible around this time due to technical difficulties. Wavelength calibration was performed using the telluric skylines in the spectra. The flux calibration used the spectroscopic standard star LTT3864, and a solar analog star, HD 43965, was used for slope correction. A combination of the DRAGONS Gemini reduction software \citep[][]{Labrie2019} and custom software was used to reduce the GMOS-S spectra. A technical defect in the amplifier of the GMOS-S instrument affected the extraction of spectra between 0.69-0.81~micron, so the data from this wavelength range were removed from the final spectrum.

\section{Results}
Fig.~1 shows the Keck/LRIS and Gemini S/GMOS-S reflectance spectra of \vdns, produced by dividing the asteroid spectrum by the solar analog spectrum and normalizing it at 0.55~micron. We do not show the data shortward of 0.45 micron or longward of 1.03 micron where the S/N becomes low. The spectrum of \vd has a normalized gradient of 13$\pm$2$\%$/ 100 nm between 0.5-0.6~micron. Equivalent visible color indices corresponding to SDSS g, r, i and z bandpasses \citep[][]{Fukugita1996} are $g$-$r$ = 0.67$\pm$0.03 and $r$-$i$ = 0.28$\pm$0.03, $i$-$z$ = 0.11$\pm$0.02. We note that there is a slope change at approximately 0.56~micron with a steeper slope shortward of this wavelength that is seen in the spectrum of some S-type asteroids and attributed to an unidentified UV absorber \citep{McFadden2001}. Longward of $\approx$0.75~micron, the reflectance of \vd starts to decrease indicating the presence of a broad absorption band that appears to peak at roughly 0.9-1.0~micron. The reflectance spectrum obtained with Gemini S/GMOS-S is also plotted and shows similar spectral features except with a shallower 1~micron band. We discuss the possible significance of this difference below.

To assign a taxonomic class for \vd, we compare the spectra of \vd with the mean Bus-DeMean asteroid spectrum database \citep{DeMeo2009}. Visual inspection quickly identifies the \vd spectrum as best fit with the nominal S- or Sq-type (see Fig.~1). We computed the $\chi^2$ statistic for our spectra compared to the interpolated mean spectra from 0.45 and 1.03 micron. We compared to mean spectra for nineteen separate spectral types and sub-types including S-complex and related types (A, S, Sq, O, Q, R, V) which share moderate to high albedos and a 0.9-1 micron absorption band, C-complex types (B, C, Cg, Cgh, D) that are characterized by low albedo and featureless spectra, X-complex and other miscellaneous types with varying albedos and spectra with reddish slopes and slight spectral absorption features (K, L, X, Xc, Xe, Xk, Xn) \citep[][]{Bus2002,DeMeo2009}. This test confirms that the Sq-type is the best fit to the LRIS spectrum. The most important characteristics when identifying matches in the Bus-DeMeo system are the absorption features, as the system is a ``feature-based'' classification.  The correspondence of the \vd absorption feature near 0.9-1.0 micron and the Sq-class is excellent, giving the highest confidence in assigning the Sq taxonomic classification.

 For completeness, we report the next lowest $\chi^2$ fits which resulted in the fit to the K- and S-type spectra with $\chi^2$ fits that are higher by factors of 3.30 and 3.33 respectively. For the GMOS-S spectrum the S-type spectrum is the best fit, followed by the K, Xe and Sq-types that have $\chi^2$ fits a factor of $\approx$10 higher. Because of the missing data from 0.7-0.8~micron in the GMOS-S spectrum, we place a higher significance on the fit to the LRIS data. C-, V-, and Q- types, three types with significantly different albedos that are present in the inner main belt, are more than an order of magnitude worse fits and can be confidently ruled out. We conclude that \vd is a typical S- or Sq-type inner main belt asteroid. The spectral range of V, S, Sq and C-type asteroids from \citep[][]{DeMeo2013} are plotted in Fig.~1.

\section{Discussion and conclusion}

The best taxonomic match occurs between \vd and the mean spectra of the either the S- or Sq-type taxonomic class as measured both by the use of the $\chi^2$ test, characteristics of the 1-micron absorption feature, and by visual inspection of the spectra in Fig.~1. S-, Sq-, and Q-type asteroids are closely related with the 1~micron absorption feature depth being deeper in Sq-types compared to S-types, although not as deep as in Q-types \citep{Bus2002,DeMeo2009}. The varying depth of the 1~micron absorption feature in S, Sq and Q-types is thought to reflect differing amounts of space weathering on the surfaces of asteroids that otherwise share ordinary chondrite compositions \citep{Binzel2015,DeMeo2023}. On small asteroids at the km-size scale, minor collisions can shake the impacted body, replenishing its surface with fresh material from below the top layer of space-weathered material \citep{DeMeo2023}. And because the S-, Sq-, and Q- types have similar albedos \citep{Binzel2004}, an exact classification along this spectrum of similar objects is not necessary for our purposes. As a final caveat, we note that using visible spectral information alone for taxonomic classification is less precise than would result from the addition of near-infrared data as well \citep{DeMeo2009}, although in the present case, we believe there is little chance of consequential misidentification.

We can also use knowledge of the orbital distribution of asteroid types as a check on our identification. As seen in \cite{Masiero2011, DeMeo2014a}, S-type and C-types (including all of the S- and C-complex subtypes) comprise the bulk of objects with diameters ranging from 5 to 20 km in the inner main belt where \vd is located. Looking in finer detail at the mix of types as a function of heliocentric radius, we see that at a semimajor axis of 2.2 AU, S-types are by far the most numerous (see Fig. 9 in \cite{DeMeo2013}). It is worth noting that the mix of spectral types at a given location in the main belt is size-dependent, therefore, there is a potential uncertainty in estimating the fraction of each spectral type at the km size scale. Nonetheless, it is very likely that the dominant spectral types remain a mix of S- and C-types at smaller sizes. The 1~micron absorption feature seen in the spectrum of \vd rules out the C-type taxonomy \citep{Bus2002}, thus an S-type identification (inluding Sq) is the {\it a priori} statistically most-likely outcome.
 
Constraints on other asteroid spectral types that constitute lower-probability candidates (and worse spectral fits) can be made on the basis of where they are localized in the main belt compared to \vdns. Xe/E-types, for example, are predominantly located in the Hungaria region of the main belt located at $a$ $\approx$ 2 au and $i$ $\approx$ 20-25$^\circ$, well separated from the location of \vd with $a$ $\approx$ 2.2 au and $i$ $\approx$ 2$^\circ$ weakening Xe/E-types as a possible match for \vd \citep{DeMeo2013, Lucas2017}. A hypothetical K-type match for \vd is contradicted by the fact that K-types are mostly found in the Eos outer main belt asteroid family \citep{Broz2013a} confined to the 7:3 mean motion resonance with Jupiter at 2.957 au \citep{Hanus2018}. Thus, based on both the orbital distribution and spectral fit, the identification of Dinkinesh as an S- or Sq-type asteroid is strongly supported.

With the identification of \vd as an S- or Sq-type asteroid, we turn to using that information to assign a probable albedo. To do this we must make the assumption that the measured albedos of asteroids with a given spectral type can be extrapolated to other asteroids of the same type. The albedos of km-scale S, Sq, and Q Near Earth asteroids have been measured and range from 0.15 - 0.41 \citep{Delbo2003,Binzel2004}. Over a size range of 0.6-1.0 km the mean albedo is 0.26 with a formal variance of $\pm$0.03 computed from a running $n=5$ box mean. To be conservative, we adopt $\mathrm{p_v} = 0.26 \pm 0.09$ as the range most likely to encompass the actual albedo of \vd. 

The final step in estimating the size of \vd is to use the measured absolute magnitude to derive an effective diameter. Using H$_V$ = 17.63$\pm$0.04 (Mottola, private communication) and the albedo range derived above results in a size range of $\mathrm{D_{eff}} = 0.67 - 0.96$ km. This diameter range is independent of lightcurve variability. Other sources contributing to the size uncertainty, such as lightcure variabilitity, are likely comparable to or larger than that arising from the estimated albedo. The illuminated area on approach by the \lc spacecraft will depend on many factors including shape, pole position, phase angle, and topographic shadowing, all of which must be studied and modeled in advance. However, the range of possible $\mathrm{D_{eff}}$ values do, however, provide the necessary starting point for planning the \lc encounter.

While it is possible that the differences in the spectra obtained with LRIS and GMOS-S are artifacts of instrumental differences or the higher airmass at which the GMOS-S observations were obtained, which can have an effect on asteroid spectra \citep[][]{Reddy2015}, we believe it is worth considering the possibility that this measured difference reflects a real variation in \vd. As noted above, the band depth of the 0.9-1~micron feature is related to the accumulation and removal of weathered material. In a non-spherical object, it is possible that the small impacts responsible for removing weathered regolith can result in uneven distributions of such material. Areas at low geopotential height might accumulate or retain more weathered surface materials than other locations on the body. Thus, it is possible that the spectrum could change as a function of the rotational phase as seen from the Earth. The \lc spacecraft could test this possibility with visible spectrophotometry and near-infrared spectra obtained over the course of the asteroid's rotation. 

When \lc encounters \vd it will be the smallest main belt asteroid ever to be visited by a spacecraft. {\it Deep Space 1} flew by the main belt asteroid (9969) Braille on 29 July 1999, passing within 26 km (although unfortunately only obtaining imagery from a distance of about 14,000 km). Braille is classified as a Q-type asteroid and is irregularly shaped, measuring approximately $2\times1\times1$ km \citep{Oberst2001}. The smallest main belt asteroid encounter with well-resolved imaging is the E-type asteroid (2867) \v{S}teins which the {\it Rosetta} spacecraft flew by in September 2008. \v{S}teins has an effective diameter of $\mathrm{D_{eff}}$ = 5.1 km \citep{Keller+2010}, making it more than two orders of magnitude larger in volume than \vdns. Similar-sized asteroids have been explored among the Near Earth Asteroids. (65803) Didymos, recently encountered by the DART mission may be the closest in size and spectral type to \vd at an effective diameter of 0.78 km and Sq-type spectrum \citep{Cheng+2018}. (101955) Bennu and (162173) Ryugu, neatly bracket the size range of \vd \citep{Lauretta2015,Watanabe2019}, but both of these objects are classified as B/C-complex spectral types \citep{Lauretta2019,Muller2017}. The \lc encounter thus provides an opportunity to compare similar-sized objects from the main belt and Near Earth populations. \lc will also provide a definitive size and shape for \vdns, providing a near-term check on the validity of the work presented here, a prospect we eagerly anticipate.

After this paper was submitted to Icarus on February 24, 2023, a manuscript by \citet[][]{deLeon2023} was submitted on February 28, 2023 with results consistent with ours for the spectral typing of \vdns.

\bibliographystyle{aasjournal}
\bibliography{scibib}

\section*{Acknowledgements}

\noindent Based on observations obtained at the international Gemini Observatory, a program of NSF's NOIRLab, which is managed by the Association of Universities for Research in Astronomy (AURA) under a cooperative agreement with the National Science Foundation on behalf of the Gemini Observatory partnership. Some of the data presented herein were obtained at the W. M. Keck Observatory, which is operated as a scientific partnership among the California Institute of Technology, the University of California and the National Aeronautics and Space Administration. We wish to recognize and acknowledge the cultural role and reverence that the summit of Maunakea has always had within the indigenous Hawaiian community. Part of the data utilized in this publication were obtained and made available by the MITHNEOS MIT-Hawaii Near-Earth Object Spectroscopic Survey. B.T.B. is supported by an appointment to the NASA Postdoctoral Program at the NASA Goddard Space Flight Center, administered by Oak Ridge Associated Universities under contract with NASA.

\subsection*{Funding}
\noindent C.F.~acknowledges support from the Heising-Simons Foundation (grant $\#$2018-0907). The MIT component of this work is supported by NASA grant 80NSSC18K0849. The IRTF is operated by the University of Hawaii under contract 80HQTR19D0030 with the National Aeronautics and Space Administration. 

\subsection*{Software}
\noindent The LPipe reduction software \citep[][]{Perley2019} used in this work is publicly available at \url{https://sites.astro.caltech.edu/~dperley/programs/lris/manual.html}. The Gemini DRAGONS data reduction software \citep[][]{Labrie2019} is available at \url{https://github.com/GeminiDRSoftware}.

\newpage
\clearpage
\begin{figure}
\centering 
\includegraphics[width=.9\linewidth]{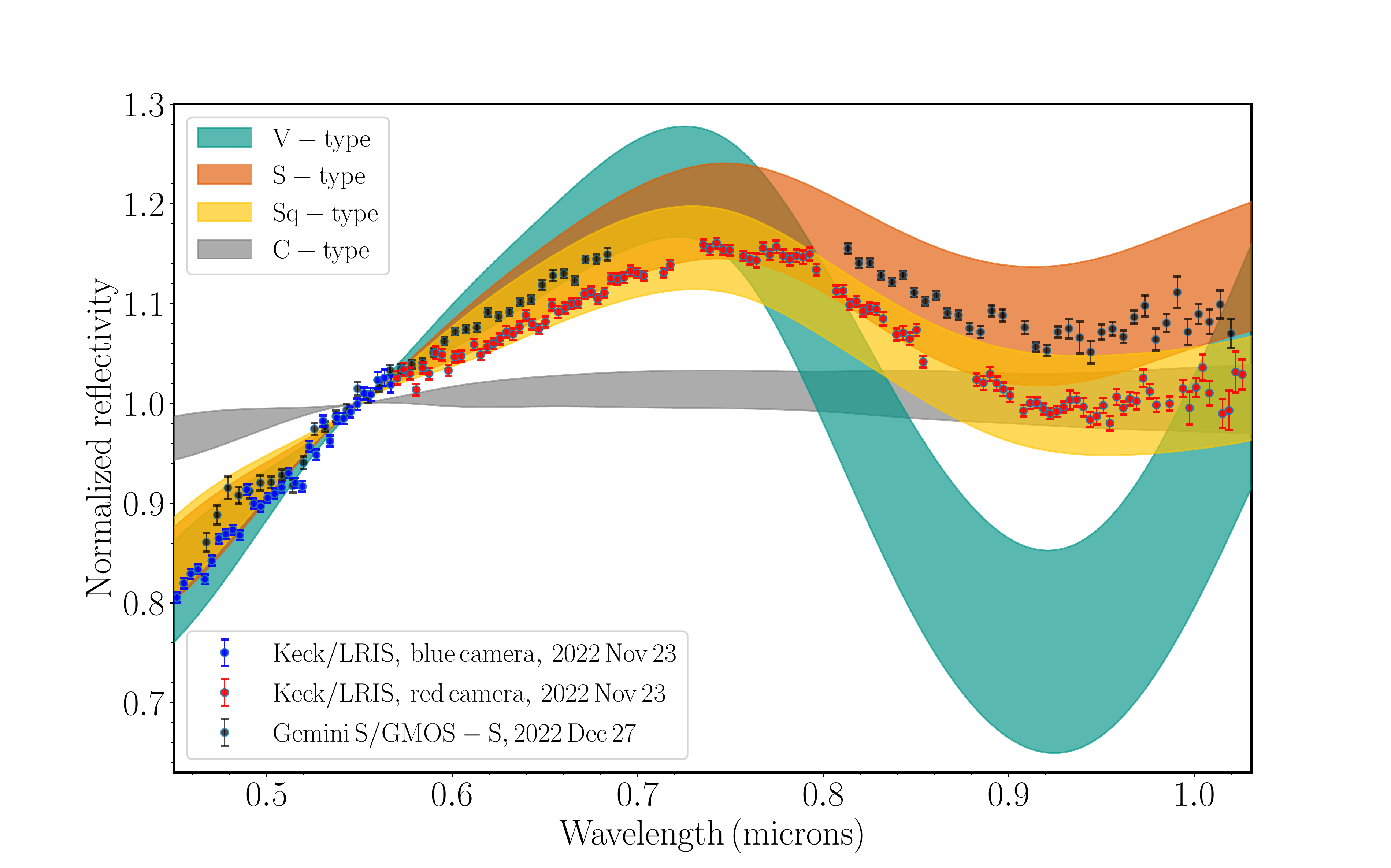}
\caption{\textbf{Visible spectra of \vd taken with Keck/LRIS and Gemini S/GMOS-S.} The spectrum of \vd taken with Keck/LRIS on 2022 November 23 is plotted in blue for the portion covered by the blue camera and plotted in red for the portion covered by the red camera.  The LRIS spectrum was obtained by combining two spectra from the blue and red camera separated by a dichroic at 0.56~micron. The bump at $\approx$0.56~micron is an artifact caused by the dichroic and the gaps at $\sim$0.72, $\sim$0.80 and $\sim$0.87~micron are artifacts and the imperfect removal of telluric features. The Gemini S/GMOS-S spectrum of \vd obtained on 2022 Dec 27 is plotted in black. The gap in the GMOS-S spectrum between 0.69-0.81 micron is due to a faulty readout amplifier on a portion of the GMOS-S detector plane. The have been normalized to unity at 0.55~micron. The LRIS and GMOS-S data have been rebinned by a factor of 30 using an error-weighted mean. The spectral ranges of V, S, Sq, and C-type asteroids from \citep{DeMeo2009} are over-plotted in aqua, red, yellow, and gray respectively.}
\end{figure}

\end{document}